\documentclass[12pt]{article}
\usepackage[pctex32]{graphics}
\makeatletter
\renewcommand{\theequation}{%
\thesection.\arabic{equation}}
\@addtoreset{equation}{section}
\newcommand{\be}{\begin{eqnarray}}
\newcommand{\ee}{\end{eqnarray}}
\newcommand{\nn}{\nonumber}

\begin{document}
\makeatletter \@addtoreset{equation}{section} \makeatother
\renewcommand{\theequation}{\thesection.\arabic{equation}}
\begin{center}
~\\{\bf  \Large  Lie 3-Algebra Non-Abelian (2,0) Theory in Loop Space} 
\vspace{3cm}

    Kuo-Wei Huang$^{\dagger 1}$,~~Wung-Hong Huang$^{*2}$\\
~\\
    $^\dagger$Department of Physics,  National  Taiwan University\\
                       Taipei, Taiwan
~\\
 $^*$Department of Physics, National Cheng Kung University\\
                       Tainan, Taiwan\\

~\\
~\\
~\\
~\\  
{\Large \bf Abstract}
\end{center}
It is believed that the multiple M5-branes are described by the non-abelian (2,0) theory and have the non-local structure.  In this note we investigate the non-abelian (2,0) theory in loop space which incorporates the non-local property.  All fields will be formulated as loop fields and the two-form potential becomes a part of connection.  We make an ansatz for field supersymmetry transformation with a help of Lie 3-algebra and examine the closure condition of the transformation to find the field equations.  However, the closure conditions lead to several complex terms and we have not yet found a simple form for some constrain field equations.  In particular, we present the clear scheme and several detailed calculations in each step.  Many useful $\Gamma$ matrix algebras are derived in the appendix. \\

\vspace{2cm}
\begin{flushleft}
$^1$E-mail:  kwhuang87@gmail.com\\
$^2$E-mail:  whhwung@mail.ncku.edu.tw\\
\end{flushleft}
\newpage
\section{Introduction}
String theories in 10D can be unified by 11D unique M-theory, where the basic objects are M2-brane and M5-brane (the magnetic version of M2-brane). One can say that M-branes are the most fundamental objects we have right now (for a review of M-theory see [1] and references within). The descriptions of single M2 or M5 brane have already known for quite a long time (for single M5-brane see [2]), while the understanding of the multiple M2 branes gained ground in the past few years by using the so-call Lie 3-algebra [3]. It is interesting to keep moving forward to start thinking about how do we describe multiple M5-branes. 
\\

We know that Maxwell field (abelian gauge theory) in a single D-brane will be extended to Yang-Mill field (non-abelian gauge theory) when D-branes start to coincide.  We also know that the strong coupling limit of 1-form potential $A^{\mu}$ in the D4-brane becomes the 2-form potential $B^{\mu\nu}$ (with self-dual 3-form field strength $H^{\mu\nu\lambda}$) in the M5-brane, thus naturally one would expect some kind of non-abelian 2-form theory will be involved in multiple M5-branes.\\

Although we expect that the multiple M5-brane to be a non-Abelian theory,  it can not be an ordinary non-Abelian gauge theory, because the entropy of coincident N M5-branes does not scale as $N^{2}$ like Yang-Mills theories but rather $N^{3}$. In the case of Multiple M2-branes  it is also not an ordinary non-Abelian theory (coincident N M2-branes scale as $N^{3\over 2}$), it has the novel gauge symmetry based on the Lie 3-algebra.  It is natural to ask oneself whether similar structure play the crucial role when one consider multiple M5-branes.
\\

On the other hand, it has been established that multiple M5-branes should be a non-local theory [4]. A natural way to deal with non-local structure is to work in the loop space, which is defined as the space of map from the circle into the manifold. A canonical analysis of the boundary of M2-brane (self-dual string) leads to a (noncommutative) loop space on the M5 brane is discussed in [5].
\\

This note is intended as an initial step of trying a possible approach by combining the concept of the Lie 3-algebra and the loop space together to see whether this method can indeed offer the key ingredients to the understanding of multiple M5-branes in the future. This note can also be considered as a attempt to link up the methods of following two papers: one recent paper using the Lie 3-algebra to construct non-abelian (2,0) theory [6] and the another paper discussed non-abelian (2,0) theory in the loop space [7]
\section{Covariant Derivative in Loop Space and Loop Field}
Let us first set up the loop space representation which we use in this note. Denote $C^{\mu}(s)$ as a function in loop space parameterized by a inner parameter $s$ and let it satisfy the loop condition : $C^{\mu}(0)=C^{\mu}(2\pi)$, where $\mu$ is ordinary space-time index. We also note that 
\be \oint ds~ \dot C^{\mu}(s)=0\ee 
in which $ \dot C_{\mu}\equiv{dC_{\mu} \over ds} $, while 
\be  \oint ds~C^{\mu}(s)\equiv 
L^{\mu}\not=0 \ee   
We define the factor $L^{\mu}$ which can be considered as the measurement of the weight when we integrate around the loop path, and that will play an important role when we modeling the SUSY transformation ansatz in the  next section.
 
 We also assume that  the loop space is noncommutative [5]    
\be [C^{\mu}(s),C^{\nu}(s)]= Q^{\mu\nu\lambda}~\dot C^{\mu}(s)\not=0. \ee 
This means noncommutativity is suggested as a replacement of usual spacetime when one uses the loop coordinates, and it is considered as the nature property of loop space itself.   It is different from the case when one need to turn on the background field, and the factor $Q^{\mu\nu\lambda}$ should be defined from the fundamental noncommutative relation of these loop coordinates.  This property is important when we try to write down the proper supersymmetry transformation ansatz in the next section. We also note that $C^{\mu}$ play the roles similar to coordinates rather than the vector field.\\

One may identify the parameter ``s" we used to represent the loop space as the same as the worldvolume parameter of  the (closed) string inside the M5 branes, but in this note we will not touch the issue of  the dynamics of the self-dual string [5].
\\

There are three kind of fields in multiple M5-branes : Scalar field $\phi ^{I}$ (I = 6,...,10), fermion field $\psi$ and the 2-form potential $B^{\mu\nu}$.
We now define the corresponding loop fields [7] 
\be  \phi ^{I}_{\mu a} (C) &\equiv& \oint ds~ \phi_{a} ^{I}(C(s)) \dot C_{\mu}(s)\\
\psi _{\mu a} (C) &\equiv& \oint ds~ \psi_{a} (C(s))  \dot C_{\mu}(s)\ee   
where a is the group index. It will be more convenient if one defines
\be  \phi ^{I}_{ a}  &\equiv& {L^{\mu}\over \sqrt{ |(L^{\nu})^{2}|}}~ \phi ^{I}_{\mu a}; \\
\psi _{ a} &\equiv& {L^{\mu}\over \sqrt{|(L^{\nu})^{2}|}}~ \psi _{\mu a} \ee   
where $\phi^I _{ a}$ has the same scale dimension as $\phi ^{I}_{\mu a}$ (so as the fermion field), and we will use $\phi^{I}_{ a}$ and $\psi _{ a}$ in the supersymmetry transformation.  In this note we also introduce a scalar field $G_{a\mu}$, as that in [6]. The corresponding loop field is 
\be  G_{a\mu}\equiv\oint ds~ G_{a} (C(s)) \dot C_{\mu}(s)\ee   
 
 Our effective gauge field (connection) is 
\be  A ^{b}_{\mu a} (C) \equiv \oint ds~ B_{\mu\nu a}^{b}(C(s))  \dot C^{\nu}(s)\ee   
thus we naturally incorporate the two-form potential B-field in the our tensor multiplet.  Note that in the strict sense, the standard gauge field which has proper gauge transformation is the one form A-field, that means in the loop representation we are not going to treat the theory as a strictly gauge theory of the two-form B-field.

The suitable covariant derivative will be
\be D_{\mu}\phi ^{I}_{a}=\partial_{\mu}\phi ^{I}_{a}-A ^{b}_{\mu a}\phi ^{I}_{b}= \partial_{\mu}\phi ^{I}_{a}- \left[\oint ds~ B_{\mu\lambda a}^{b}(C(s))  \dot C^{\lambda}(s)\right] \phi ^{I}_{b}\ee 
where the derivative in loop space is defined by
 \be \partial_{\mu}= \oint ds~ {\delta \over\delta C^{\mu}(s)}\ee 
Finally, the field strength (curvature) 
\be F_{\mu\nu a}^{b}(C)= [D_{\mu},D_{\nu}]^{b}_{~a}=\partial_{\mu} A ^{b}_{\nu a}-\partial_{\nu} A ^{b}_{\mu a}+[A_{\mu}, A_{\nu}]^{b}_{~a}\ee 
is defined by the commutator of the effective gauge field (connection).


\section{Supersymmetry Transformation}

We now start to consider the SUSY transformation. There are two important hints when we try to give the ansatz.  The first one is the SUSY transformation of multiple D4-branes, which is expected to appear when one makes a reduction on a circle for multiple M5-branes:
\be  \delta\phi^{I}&=&i\bar\epsilon \Gamma^{I}\psi\\
\delta\psi&=&\Gamma^{\mu}\Gamma^{I}D_{\mu}\phi^{I} \epsilon+{1\over 2}\Gamma^{\mu\nu}\Gamma^{5}F_{\mu\nu}\epsilon-{i\over2}[\phi^{I},\phi^{J}]\Gamma^{IJ}\epsilon\\
\delta A_{\mu}&=&i\bar\epsilon\Gamma_{\mu}\Gamma^{5}\psi\ee   
where $\mu=(0,...,4)$; $I=(5,...,9)$ and $F_{\mu\nu}$ is field strength of Maxwell field. This can be obtained by the dimensional reduction of 10D Yang-Mill theory. 
\\

The second hint is the SUSY transformation of the single M5-brane [2]
\be  \delta\phi^{I}&=&i\bar\epsilon \Gamma^{I}\psi\\
\delta\psi&=&\Gamma^{\mu}\Gamma^{I}\partial_{\mu}\phi^{I} \epsilon+{1\over 12}\Gamma^{\mu\nu\lambda}H_{\mu\nu\lambda}\epsilon\\
\delta B_{\mu\nu}&=&i\bar\epsilon\Gamma_{\mu\nu}\psi\ee   
where $\mu=(0,...,5)$; $I=(6,....10)$ and $H_{\mu\nu\lambda}=3\partial_{[\mu}B_{\nu\lambda]}$ is self-dual 3-form. Note that the supersymmetry parameter $\epsilon$ and $\psi$ have opposite chirality due to the fact that all the fermions are Goldstinos which correspond to broken supersymmetry while the parameter correspond to unbroken supersymmetry. We take the convention: $\Gamma ^{012345}\epsilon=+ \epsilon$ and $\Gamma ^{012345}\psi=- \psi$.
\\

With the correct chirality in mind, in this note we consider the following SUSY transformation ansatz of non-abelian (2,0) theory via three-algebra in loop space
\\
\be  \delta\phi^{I}_{a}&=&i\bar\epsilon \Gamma^{I}\psi_{ a}\\
\delta\psi_{a}&=&\Gamma^{\mu}\Gamma^{I}D_{\mu}\phi^{I}_{a} \epsilon
+\kappa\Gamma^{IJK}\Gamma^{\mu\nu\lambda} Q_{\mu\nu\lambda}\phi^{I}_{b}\phi^{J}_{c}\phi^{K}_{ d} f^{cdb}_{~~~a} \epsilon
+ \Gamma^{\mu\nu\lambda}F_{\mu\nu bc}G_{d\lambda}f^{bcd}_{~~~a}\epsilon\\
\delta A ^{b}_{\mu a}&=&i\bar\epsilon\Gamma_{\mu\nu\lambda\sigma}\Gamma^{I}Q^{\nu\lambda\sigma}\psi_{c}\phi^{I}_{d}f^{cdb}_{~~~a}\\  
\delta G _{a\mu}&=&0\ee 
where $f^{cdb}_{~~~a}$ is the structure constant of Lie 3-algebra. Note that we have 4 fields in the theory while have 5 terms in supersymmetry transformation, and one  coefficients $\kappa$ should be determined by  the closure of supersymmetry transformation. \\

To proceed, let us first make some remarks about the SUSY transformation ansatz. \\

$\bullet$  It is especially noteworthy that if we want an ansatz which can indeed close on shell, we have to first take a look at the Fierz identity (see Appendix) to insure that the transformed terms have the proper form for each others so that we can extract out the equation of motions. 
\\

$\bullet$ The total anti-symmetry  gamma matrix $\Gamma^{\mu\nu\lambda}$ in front of the second term of the fermion field SUSY transformation implies that the factor $Q^{\mu\nu\lambda}$ is also total anti-symmetry, just like $Q^{\mu\nu\lambda}\sim \epsilon^{\mu\nu\lambda}$ in [5].
\\

$\bullet$ In order to find the connection to multiple D4-branes, we have to consider dimensional reduction, besides we also need to consider the process of going back to the ordinary local theory where all fields decouple from the loop. In this sense we can take the limit when the loop shrink into an point and consider the Q factor reduce to a normalized constant. 
\\

$\bullet$ Above ansatz for the SUSY transformation have the following consistent scaling dimensions   
\be  [\phi^{I}_{a}]=1~,~[\psi_{a}]={3\over 2}~,~[A ^{b}_{\mu a}]=1~,~[\epsilon]=-{1\over2}~,~ [C^{\mu}]=-1\nn  \ee   
\be  [F^{\mu\nu b}_{~~~~a}]=2~,~[Q^{\mu\nu\lambda}]=-1 ~,~ [\partial^{\mu}]=[D^{\mu}]=1~,~ [G_{a\mu}]=[s]=0
\ee   
and chiral property discussed in appendix A.
\section{Closures of Supersymmetry Transformation : Scalar Field}
Now, we start to examine our ansatz by the closure of superalgebra. First, let us check our (loop) scalar field. Use the ansatz we have
\be 
\delta_1\delta_2\phi^I_{a}=i\bar\epsilon_2 \Gamma^{I}(\delta_1\psi_{a})=i\bar\epsilon_2 \Gamma^{I}\Big(\Gamma^{\mu}\Gamma^{\bar I}D_{\mu}\phi^{\bar I}_{a} \epsilon_1 &+&\kappa\Gamma^{\bar I\bar J\bar K}\Gamma^{\mu\nu\lambda} Q_{\mu\nu\lambda}\phi^{\bar I}_{b}\phi^{\bar J}_{c}\phi^{\bar K}_{ d} f^{bcd}_{~~~a} \epsilon_1
\nn\\
&+&\Gamma^{\mu\nu\lambda}F_{\mu\nu bc}G_{d\lambda}f^{bcd}_{~~~a}\epsilon_1\Big)
\ee 
and we have to  deal with  three terms (denoted as $\phi1$, $\phi 2$ and $\phi 3$) in above equation.   
\\

$\bullet $ $\phi 1$ term: Use the $\Gamma$ matrix property derived in appendex we know that
\be
\Gamma^{I}\Gamma^{\mu}\Gamma^{\bar I}&=&-\Gamma^{I\bar I}\Gamma^{\mu}-g^{I\bar I}\Gamma^{\mu}\nn\\
&\ddot=& -g^{I\bar I}\Gamma^{\mu}\ee 
We use $\ddot=$ to denote that we have dropped the terms which become zero in  considering the commutator $[\delta_1,\delta_2]$.  Which terms shall be dropped could be easily determined  from the symmetry property proved in the appendex A.  Therefore, $\phi 1$ term will contribute following result to  $[\delta_{1},\delta_{2}]\phi^{I}_a$ :
\be [\delta_{1},\delta_{2}]\phi^{I}_{a_{|\phi 1}}=(-2i\bar\epsilon_{2}\Gamma^{\mu}\epsilon_{1})D_{\mu}\phi^{I}_{a} \ee 

$\bullet $ $\phi 2$ term: Use the $\Gamma$ matrix property derived in appendex we know that
\be\Gamma^{I}\Gamma^{\bar I\bar J\bar K}\Gamma^{\mu\nu\lambda}&=&(\Gamma^{I\bar I\bar J\bar K}+ 3g^{I[\bar I}\Gamma^{\bar J\bar K]})\Gamma^{\mu\nu\lambda}~\ddot = ~3g^{I[\bar I}\Gamma^{\bar J\bar K]}\Gamma^{\mu\nu\lambda}\ee 
Therefore, this term will contribute following result to  $[\delta_{1},\delta_{2}]\phi^{I}_{ a}$ :
\be [\delta_{1},\delta_{2}]\phi^{I}_{a_{|\phi 2}}=({i 6 \kappa}\bar\epsilon_{2}\Gamma_{\mu\nu\lambda}\Gamma^{KL}\epsilon_{1})Q^{\mu\nu\lambda}\phi^{K}_{c}\phi^{L}_{d}f^{cdb}_{~~~a}\phi^{I}_{b}\ee

$\bullet $ $\phi 3$ term: Use the $\Gamma$ matrix property derived in appendex we know that
\be\Gamma^{I}\Gamma^{\mu\nu\lambda}&\ddot= &0 
\ee
Therefore, this term  contribute null to  $[\delta_{1},\delta_{2}]\phi^{I}_{ a}$.
\\

Thus we define the translation parameter $\upsilon^{\mu}$ and  gauge transformation parameter $\Lambda^{b}_{~a}$ by
\be
v^{\mu} &\equiv& -2i\bar\epsilon_{2}\Gamma^{\mu}\epsilon_{1} \\
\Lambda^{b}_{~a}&\equiv&{i 6 \kappa}\bar\epsilon_{2}\Gamma_{\mu\nu\lambda}\Gamma^{KL}\epsilon_{1}Q^{\mu\nu\lambda}\phi^{K}_{c}\phi^{L}_{d}f^{cdb}_{~~~a}
\ee
and find that 
\be  [\delta_{1},\delta_{2}]\phi^{I}_{ a} &=& (-2i\bar\epsilon_{2}\Gamma^{\mu}\epsilon_{1})D_{\mu}\phi^{I}_{a}+({i 6 \kappa}\bar\epsilon_{2}\Gamma_{\mu\nu\lambda}\Gamma^{KL}\epsilon_{1})Q^{\mu\nu\lambda}\phi^{K}_{c}\phi^{L}_{d}f^{cdb}_{~~~a}\phi^{I}_{b}\nn\\
 &\equiv& v^{\mu} D_{\mu}\phi^{I}_{a}+\Lambda^{b}_{~a} \phi^{I}_{ b}\ee   
which tells us that the supersymmetry transformation closes into a translation and a gauge transformation.
\section{Closures of Supersymmetry Transformation : Gauge Field}
Next we turn to our gauge field, this term will give us the field equation of field strength.  

First, use the ansatz we see that
\be
\delta_1\delta_2 A ^{b}_{\mu a}&=&i\bar\epsilon_2\Gamma_{\mu\nu\lambda\sigma}\Gamma^{I}Q^{\nu\lambda\sigma}~\delta_1(\psi_{c} \cdot \phi^{I}_{d})f^{cdb}_{~~~a}=i\bar\epsilon_2\Gamma_{\mu\nu\lambda\sigma}\Gamma^{I}Q^{\nu\lambda\sigma}(\delta_1\psi_{c} \cdot \phi^{I}_{d}+\psi_{c} \cdot \delta_1\phi^{I}_{d})f^{cdb}_{~~~a}\nn\\
&=&i\bar\epsilon_2\Gamma_{\mu\nu\lambda\sigma}\Gamma^{I}Q^{\nu\lambda\sigma}\Big(\Gamma^{\bar \mu}\Gamma^{\bar I}D_{\bar \mu}\phi^{\bar I}_{c} \epsilon_1 +\kappa\Gamma^{\bar I\bar J\bar K}\Gamma^{\bar \mu\bar \nu\bar \lambda} Q_{\bar \mu\bar \nu\bar \lambda}\phi^{\bar I}_{\bar b}\phi^{\bar J}_{\bar c}\phi^{\bar K}_{ \bar d} f^{\bar c\bar d\bar b}_{~~~c} \epsilon_1
\nn\\
&&+ \Gamma^{\bar \mu\bar \nu\bar \lambda}F_{\bar \mu\bar \nu \bar b\bar c}G_{\bar d\bar \lambda}\epsilon_1~f^{\bar b \bar c \bar d}_{~~~c}\Big) \cdot \phi^{I}_{d}~f^{bcd}_{~~~a}+i\bar\epsilon_2\Gamma_{\mu\nu\lambda\sigma}\Gamma^{I}Q^{\nu\lambda\sigma}(\psi_{c} \cdot i\bar\epsilon_1\Gamma^{I}\psi_{d})f^{cdb}_{~~~a}
\ee
and we have to  deal with  four terms (denoted as $A 1$, $A 2$, $A 3$ and $A 4$) in above equation.   
\\

$\bullet $ $A1$ term: Use the $\Gamma$ matrix property derived in appendex we know that
\be
\Gamma^I\Gamma_{\mu\nu\lambda\rho}\Gamma^{\bar\mu}\Gamma^{\bar I}&=&-\Gamma_{\mu\nu\lambda\rho}\Gamma^{\bar\mu}\Gamma^I\Gamma^{\bar I}=-\Big(\Gamma_{\mu\nu\lambda\rho}^{~~~~~\bar\mu}+4\Gamma_{[\mu\nu\lambda}g_{\rho]}^{\bar\mu} \Big)\Big(\Gamma^{I\bar I}+g^{I\bar I}\Big)\nn\\
&\ddot=&-\Gamma_{\mu\nu\lambda\rho}^{~~~~~\bar\mu}g^{I\bar I}-4\Gamma_{[\mu\nu\lambda}g_{\rho]^{\bar\mu}}\Gamma^{I\bar I}
\ee
Therefore, this term will contribute following result to  $[\delta_{1},\delta_{2}]A^{b}_{\mu a}$ :
\be [\delta_{1},\delta_{2}]A^{b}_{\mu a_{|{A1}}}=\upsilon^{\delta}\epsilon_{\mu\nu\lambda\rho\sigma\delta}\Big(Q^{\nu\lambda\rho}\phi_{c}^I D^\sigma \phi_d^I f^{cdb}_{~~~a}  \Big)+{2\over 3 \kappa}D_\mu \Lambda^b_a \ee

$\bullet $ $A2$  term:  This term is proportional to 
\be
[\delta_{1},\delta_{2}]A^{b}_{\mu a_{|{A2}}}=\epsilon_2\Gamma_{\mu\nu\lambda\sigma}\Gamma^{I}\Gamma^{\bar I\bar J\bar K}\Gamma^{\nu\lambda\rho}\epsilon_1 \phi^{\bar I}_{\bar b}\phi^{\bar J}_{\bar c}\phi^{\bar K}_{ \bar d}\phi^{I}_{d}~f^{\bar c\bar d\bar b}_{~~~c}f^{bcd}_{~~~a}\ee 
which vanishes as a consequence of fundamental identity of Lie 3-algebra, as analogic term in BLG multiple M2 theory [3].
\\

$\bullet $ $A3$ term: Use the $\Gamma$ matrix property derived in appendex we know that
\be  \Gamma_{\mu\nu\lambda\rho} \Gamma^{\bar\mu\bar\nu\bar\lambda}&=&\Gamma_{\mu\nu\lambda\rho}^{~~~~~\bar\mu\bar\nu\bar\lambda}
+12\Gamma_{[\mu\nu\lambda}^{~~~~[\bar\nu\bar\lambda}g_{\rho]}^{\bar\mu]}+36\Gamma_{[\mu\nu}^{~~~[\bar\lambda}g_{\lambda\rho]}^{\bar\mu\bar\nu]}+24\Gamma_{[\mu}g_{\nu\lambda\rho]}^{\bar\mu\bar\nu\bar\lambda]}\nn\\
&\ddot=&12\Gamma_{[\mu\nu\lambda}^{~~~~[\bar\nu\bar\lambda}g_{\rho]}^{\bar\mu]}+24\Gamma_{[\mu}g_{\nu\lambda\rho]}^{\bar\mu\bar\nu\bar\lambda]}
\ee  
Therefore, this term will contribute following result to  $[\delta_{1},\delta_{2}]A^{I}_{ a}$ :
\be [\delta_{1},\delta_{2}]A^{b}_{\mu a_{|{A3}}}=i  (\bar\epsilon_{2}\Gamma^{I}\Gamma_{\sigma}\epsilon_{1}) \Big(12\epsilon_{\mu\nu\lambda}^{~~~~\bar\nu\bar\lambda\sigma}Q^{\nu\lambda\rho} F_{\rho\bar\nu\bar b\bar c}G_{\bar d\bar\lambda}+ 24 Q^{\nu\lambda\sigma} F_{\mu\nu\bar b\bar c}G_{\bar d\lambda}\Big)\phi_c^I~f^{\bar b\bar c\bar d}_{~~~d}f^{cdb}_{~~~a}\ee

$\bullet $ $A4$ term: Use the Fierz identities derived in appendex we see that  the $A$-4 term becomes
\be [\delta_{1},\delta_{2}]A^{b}_{\mu a_{|{A4}}}&=& i\bar\epsilon_2\Gamma_{\mu\nu\lambda\sigma}\Gamma^{I}Q^{\nu\lambda\sigma}(\psi_{d} \cdot i\bar\epsilon_1\Gamma^{I}\psi_{c})f^{cdb}_{~~~a}\nn\\
&=&i\bar\epsilon_2\Gamma_{\mu\nu\lambda\sigma}\Gamma^{I}Q^{\nu\lambda\sigma}(i\bar\psi_{c}\Gamma^{I}\epsilon_1)\psi_{d}f^{cdb}_{~~~a}\nn\\
&=&-{1\over 16}\bar\epsilon_2\Gamma_{\mu\nu\lambda\sigma}\Gamma^{I}\Big[~2(\bar \psi_{c}\Gamma_{\bar\mu}\psi_{d})\Gamma^{\bar\mu}\Gamma^{I}\epsilon_1\nn\\
&&-2(\bar \psi_{c}\Gamma_{\bar\mu}\Gamma^{\bar I}\epsilon_{1})\Gamma^{\bar\mu}\Gamma^{\bar I}\Gamma^{I}\psi_{d} \nn\\
 &&+{1\over 12}(\bar \psi_{c}\Gamma_{\bar\mu\bar\nu\bar\lambda}\Gamma^{\bar I \bar J}\psi_{d})\Gamma^{\bar\mu\bar\nu\bar\lambda}\Gamma^{\bar I \bar J}\Gamma^{I}\epsilon_1\Big]Q^{\nu\lambda\sigma}f^{cdb}_{~~~a}
\ee
and we have to  deal with  three terms (denoted as $A4a$, $A4b$ and $A4c$) in above equation.   
\\

$~~~\circ$ $A4a$:  Use $\Gamma$ matrix property
\be \Gamma^I\Gamma_{\mu\nu\lambda\rho}\Gamma^{\bar\mu}\Gamma^I &=&-\Gamma^I\Gamma^I\Gamma_{\mu\nu\lambda\rho}\Gamma^{\bar\mu}=-5\Big(\Gamma_{\mu\nu\lambda\rho}^{\bar\mu}+4 \Gamma_{[\mu\nu\lambda}g_{\rho]}^{\bar\mu}\Big)\nn\\
&\ddot =& -5\Gamma_{\mu\nu\lambda\rho}^{~~~~~\bar\mu}
\ee
 we find the contribution
\be
 [\delta_{1},\delta_{2}]A^{b}_{\mu a_{|{A4a}}}&=&{5\over 16}(-i2\bar\epsilon_2\Gamma_{\mu\nu\lambda\rho}^{~~~~~\bar\mu} \epsilon_1)(i\bar \psi_{c}\Gamma_{\bar\mu}\psi_{d})Q^{\nu\lambda\sigma}f^{cdb}_{~~~a}\nn\\
&=&{5\over 16}\upsilon^\sigma \varepsilon_{\mu\nu\lambda\rho\delta\sigma}(i\bar \psi_{c}\Gamma^{\delta}\psi_{d})Q^{\nu\lambda\sigma}f^{cdb}_{~~~a}
\ee
\\

$~~~\circ$ $A4b$:  Use $\Gamma$ matrix property
\be \Gamma^I\Gamma_{\mu\nu\lambda\rho}\Gamma^{\bar\mu}\Gamma^{\bar I}\Gamma^I&=&-\Gamma_{\mu\nu\lambda\rho}\Gamma^{\bar\mu}~\Big(\Gamma^{I}\Gamma^{\bar I}\Gamma^{I}\Big)=3\Big(\Gamma_{\mu\nu\lambda\rho}^{~~~~\bar\mu}+4 \Gamma_{[\mu\nu\lambda}g_{\rho}^{\bar\mu}\Big)\Gamma^{\bar I}\nn\\
 &\ddot =& 3\Gamma_{\mu\nu\lambda\rho}^{~~~~~\bar\mu}\Gamma^{\bar I}
\ee
 we find the contribution
\be
 [\delta_{1},\delta_{2}]A^{b}_{\mu a_{|{A4b}}}&=&{3\over 8}(\bar\epsilon_2\Gamma_{\mu\nu\lambda\rho}^{~~~~~\bar\mu}\Gamma^{\bar I} \epsilon_1)(\bar \psi_{c}\Gamma_{\bar\mu}\Gamma^{\bar I}\psi_{d})Q^{\nu\lambda\sigma}f^{cdb}_{~~~a}\nn\\
&=&{3\over 8}(\bar\epsilon_2\Gamma^{\sigma}\Gamma^{\bar I} \epsilon_1)\varepsilon_{\mu\nu\lambda\rho\delta\sigma}(\bar \psi_{c}\Gamma^{\delta}\Gamma^{\bar I} \psi_{d})Q^{\nu\lambda\sigma}f^{cdb}_{~~~a}
\ee

$~~~\circ$ $A4c$:  Use $\Gamma$ matrix property
\be  \Gamma_{\mu\nu\lambda\rho} \Gamma^{\bar\mu\bar\nu\bar\lambda}&=&\Gamma_{\mu\nu\lambda\rho}^{~~~~~\bar\mu\bar\nu\bar\lambda}
+12\Gamma_{[\mu\nu\lambda}^{~~~~[\bar\nu\bar\lambda}g_{\rho]}^{\bar\mu]}+36\Gamma_{[\mu\nu}^{~~~[\bar\lambda}g_{\lambda\rho]}^{\bar\mu\bar\nu]}+24\Gamma_{[\mu}g_{\nu\lambda\rho]}^{\bar\mu\bar\nu\bar\lambda]}\nn\\
\Gamma^{I}\Gamma^{\bar I\bar J}\Gamma^{J}&=& \Gamma^{\bar I\bar J}
\ee
We find that this term  contribute null to  $[\delta_{1},\delta_{2}]A^{b}_{\mu a}$.
\\

So we find that 
\be  [\delta_{1},\delta_{2}]A ^{b}_{\mu a}&=& i
\upsilon^\delta \varepsilon_{\mu\nu\lambda\rho\delta\sigma}\Big(\phi_c^I D^\sigma \phi_d^I + i {5\over 16}\bar \psi_{c}\Gamma^{\sigma}\psi_{d}\Big) Q^{\nu\lambda\rho}f^{cdb}_{~~~a}+{2\over3\kappa} D_\mu\Lambda_a^b\nn\\
&&+i(\bar\epsilon_2\Gamma^{I}\Gamma_{\sigma} \epsilon_1)\Big( 12 ( \epsilon_{\mu\nu\lambda}^{~~~~\bar\nu\bar\lambda\sigma}Q^{\nu\lambda\rho} F_{\rho\bar\nu\bar b\bar c}G_{\bar d\bar\lambda}+ 2  Q^{\nu\lambda\sigma} F_{\mu\nu\bar b\bar c}G_{\bar d\lambda})\phi_c^I~f^{\bar b\bar c\bar d}_{~~~d}\nn\\
&&-{3\over 8}\varepsilon_{\mu\nu\lambda\rho\delta}^{~~~~~~\sigma}(\bar \psi_{c}\Gamma^{\delta}\Gamma^{I} \psi_{d})Q^{\nu\lambda\rho}\Big)f^{cdb}_{~~~a}\nn\\
&\equiv& \upsilon^\delta F_{\mu\delta a}^b+ D_\mu\Lambda_a^b\ee   
 Thus the supersymmetry transformation closes into a translation and a gauge transformation.  This implies that  
\be \kappa={2\over3}\ee
and the gauge field equation is 
\be  F_{\mu\nu a}^b-\varepsilon_{\mu\nu\lambda\rho\delta\sigma}\Big(\phi_c^I D^\sigma \phi_d^I + i {5\over 16}\bar \psi_{c}\Gamma^{\sigma}\psi_{d}\Big) Q^{\rho\lambda\delta}f^{cdb}_{~~~a}=0
\ee   
Besides, we also have another constrain field equation
\be
12 ( \epsilon_{\mu\nu\lambda}^{~~~~\bar\nu\bar\lambda\sigma}Q^{\nu\lambda\rho} F_{\rho\bar\nu\bar b\bar c}G_{\bar d\bar\lambda}+ 2  Q^{\nu\lambda\sigma} F_{\mu\nu\bar b\bar c}G_{\bar d\lambda})\phi_c^I~f^{\bar b\bar c\bar d}_{~~~d}-{3\over 8}\varepsilon_{\mu\nu\lambda\rho\delta}^{~~~~~~\sigma}Q^{\nu\lambda\rho}(\bar \psi_{c}\Gamma^{\delta}\Gamma^{I} \psi_{d})=0\nn\\
\ee
It is hoped that the constrain field equation may be simplified (or automatically be satisfied) from the property of closures of  fermion field supersymmetry transformation analyzed  in next section.
\section{Closures of Supersymmetry Transformation : Fermion Field}
Finally, we deal with the closure of fermion term, this term will give us the fermion field equation.  

Use the ansatz we see that
\be
\delta_1\delta_2\psi_{a}&=&\Gamma^{\mu}\Gamma^{I}D_{\mu}(\delta_1\phi^{I}_{a}) \epsilon_2-\Gamma^{\mu}\Gamma^{I}(\delta_1 A_{\mu a}^b)\phi^{I}_{b} \epsilon_2\nn\\
&&+3 \kappa\Gamma^{IJK}\Gamma^{\mu\nu\lambda} Q_{\mu\nu\lambda}(\delta_1\phi^{I}_{b})\phi^{J}_{c}\phi^{K}_{ d} f^{cdb}_{~~~a} \epsilon_2
+ \Gamma^{\mu\nu\lambda}(\delta_1F_{\mu\nu bc})G_{d\lambda}f^{bcd}_{~~~a}\epsilon_2
\ee 
and we have to  deal with  four terms (denoted as $\psi 1$, $\psi A$, $\psi 2$ and $\psi F$) in above equation. 
\\

$\bullet $ Let us first consider the $\psi 1$ term.  From 
\be
\delta_1\delta_2\psi_{a}|_{\psi 1}=\Gamma^{\mu}\Gamma^{I}D_{\mu}(\delta_1\phi^{I}_{a}) \epsilon_2=\Gamma^{\mu}\Gamma^{I}D_{\mu}(i \bar\epsilon_1\Gamma^{I}\psi_{a}) \epsilon_2
\ee
and Fierz identities, the $\psi 1$ term contributes following to $[\delta_1,\delta_2]\psi_{a}$
\be
[\delta_1,\delta_2]\psi_{a|_{\psi 1}}={i\over16}\Gamma^{\mu}\Gamma^{I}\Big[2(\bar\epsilon_{2}\Gamma_{\bar\mu}\epsilon_{1})\Gamma^{\bar\mu}\Gamma^{I}D_{\mu}\psi_{a}-2(\bar\epsilon_{2}\Gamma_{\bar\mu}\Gamma^{\bar I}\epsilon_{1})\Gamma^{\bar\mu}\Gamma^{\bar I}\Gamma^{I}D_{\mu}\psi_{a}\nn \\
 +{1\over 12}(\bar\epsilon_{2}\Gamma_{\bar\mu\bar\nu\bar\lambda}\Gamma^{\bar I\bar J}\epsilon_{1})\Gamma^{\bar\mu\bar\nu\bar\lambda}\Gamma^{\bar I \bar J}\Gamma^{I}D_{\mu}\psi_{a}\Big]
\ee
Use the $\Gamma$ matrix property
\be \Gamma^\mu \Gamma^I\Gamma^{\bar \mu}\Gamma^I&=&-\Gamma^\mu \Gamma^{\bar \mu}\Gamma^I\Gamma^I = -5\Big(2g^{\mu \bar \mu} -\Gamma^{\bar \mu}\Gamma^\mu \Big)\\
\Gamma^\mu \Gamma^I\Gamma^{\bar \mu}\Gamma^{\bar I}\Gamma^I&=&-\Gamma^\mu \Gamma^{\bar \mu}\Gamma^I\Gamma^{\bar I}\Gamma^I = 3\Big(2g^{\mu \bar \mu}- \Gamma^{\bar \mu}\Gamma^\mu \Big)\Gamma^{\bar I}\\
\Gamma^\mu \Gamma^I\Gamma^{\bar \mu\bar \nu\bar \lambda}\Gamma^{\bar I\bar J}\Gamma^I&=&-\Gamma^\mu \Gamma^{\bar \mu\bar \nu\bar \lambda}\Gamma^I\Gamma^{\bar I\bar J}\Gamma^I = -\Big(6 g^{\mu [\bar \mu}\Gamma^{\bar \nu\bar \lambda]}- \Gamma^{\bar \mu\bar \nu\bar \lambda}\Gamma^\mu \Big)\Gamma^{\bar I\bar J}
\ee
we find the contributions
\be
 [\delta_1,\delta_2]\psi_{a|_{\psi 1}}&=&i(\bar\epsilon_{2}\Gamma_{\bar\mu}\epsilon_{1})\Big[-{5\over4}D^{\bar\mu}\psi_{a} +{5\over 8} \Gamma^{\bar\mu}\Gamma^{\mu}D_{\mu}\psi_{a}\Big]\nn\\
&&+i(\bar\epsilon_{2}\Gamma_{\bar\mu}\Gamma^{\bar I}\epsilon_{1})[{-3\over4}\Gamma^{\bar I}D^{\bar\mu}\psi_{a} -{3\over8} \Gamma^{\bar\mu}\Gamma^{\bar I}\Gamma^{\mu}D_{\mu}\psi_{a}\Big]\nn \\
&& +{i\over 192}(\bar\epsilon_{2}\Gamma_{\bar\mu\bar\nu\bar\lambda}\Gamma^{\bar I\bar J}\epsilon_{1})\Big[-6g^{\mu[\bar\mu}\Gamma^{\bar\nu\bar\lambda]}\Gamma^{\bar I\bar J}D_{\mu}\psi_{a}+\Gamma^{\bar\mu\bar\nu\bar\lambda}\Gamma^{\bar I \bar J}\Gamma^{\mu}D_{\mu}\psi_{a}\Big]
\ee
\\

$\bullet $  In the same way,  using  $\Gamma$ matrix properties derived in appendix it is straightforward (while sightly complex)  to calculate other terms.  We finally find that
\be
[\delta_1,\delta_2]\psi_{a}= i(\bar\epsilon_{2}\Gamma_{\mu}\epsilon_{1})~&\times&~\Big[-{5\over4}D^{\mu}\psi_{a} +{27\over4} \kappa~ \Gamma_{\nu\lambda} Q^{\mu\nu\lambda}\Gamma^{JK}\psi_b \phi^{J}_{c}\phi^{K}_{d} f^{bcd}_{~~~a}\nn\\
&~&+{3\over4} ~ \Gamma_{\nu\lambda} Q^{\mu\nu\lambda}\Gamma^{JK}\psi_b \phi^{J}_{c}\phi^{K}_{d} f^{bcd}_{~~~a}\nn\\
&~&+{3\over2}\Gamma^I \Gamma^{\nu\lambda}\Gamma_{\nu\bar\mu\bar\nu\bar\lambda} Q^{\bar\mu\bar\nu\bar\lambda}(\psi_{\bar d}D^\mu\phi^I_{\bar c}+\phi^I_{\bar c} D^\mu\psi_{\bar d} )G_{d\lambda} f^{\bar c\bar d}_{~~bc}f^{bcd}_{~~~a}\Big]\nn\\
 + i(\bar\epsilon_{2}\Gamma_{\bar\mu}\epsilon_{1})\Gamma^{\bar\mu}~&\times&~\Big[{5\over8}\Gamma^{\mu}D_{\mu}\psi_{a} -{9\over8} \kappa~ \Gamma_{\mu\nu\lambda} Q^{\mu\nu\lambda}\Gamma^{JK}\psi_b \phi^{J}_{c}\phi^{K}_{d} f^{bcd}_{~~~a}\nn\\
&~&+{5\over8} ~ \Gamma_{\mu\nu\lambda} Q^{\mu\nu\lambda}\Gamma^{JK}\psi_b \phi^{J}_{c}\phi^{K}_{d} f^{bcd}_{~~~a}\nn\\
&~&+{1\over4}\Gamma^I \Gamma^{\mu\nu\lambda}\Gamma_{\nu\bar\nu\bar\lambda\bar\rho} Q^{\bar\mu\bar\nu\bar\lambda}(\psi_{\bar d}D_\mu\phi^I_{\bar c}+\phi^I_{\bar c} D_\mu\psi_{\bar d} )G_{d\lambda} f^{\bar c\bar d}_{~~bc}f^{bcd}_{~~~a}\Big]\nn\\
+i(\bar\epsilon_{2}\Gamma_{\bar\mu}\Gamma^{\bar I}\epsilon_{1})~&\times&~\Big[{-3\over4}\Gamma^{\bar I}D^{\bar\mu}\psi_{a}+\cdot\cdot\cdot\cdot\cdot\cdot\cdot\cdot\cdot\cdot\cdot\cdot\Big]\nn \\
 +i(\bar\epsilon_{2}\Gamma_{\bar\mu}\Gamma^{\bar I}\epsilon_{1})\Gamma^{\bar\mu}\Gamma^{\bar I}~&\times&~\Big[ -{3\over8}\Gamma^{\mu} D_{\mu}\psi_{a}+\cdot\cdot\cdot\cdot\cdot\cdot\cdot\cdot\cdot\cdot\cdot\Big]\nn\\
+i(\bar\epsilon_{2}\Gamma_{\bar\mu\bar\nu\bar\lambda}\Gamma^{\bar I\bar J}\epsilon_{1})~&\times&~\Big[{-6\over192}g^{\mu[\bar\mu}\Gamma^{\bar\nu\bar\lambda]}\Gamma^{\bar I\bar J}D_{\mu}\psi_{a}+\cdot\cdot\cdot\cdot\Big]\nn\\
+ i(\bar\epsilon_{2}\Gamma_{\bar\mu\bar\nu\bar\lambda}\Gamma^{\bar I\bar J}\epsilon_{1})\Gamma^{\bar\mu\bar\nu\bar\lambda}\Gamma^{\bar I\bar J}~&\times&~\Big[{1\over192}\Gamma^{\mu}D_{\mu}\psi_{a}+\cdot\cdot\cdot\cdot\cdot\cdot\cdot\cdot\cdot\cdot\cdot~\Big]
\ee
in which the first, second, third and forth terms within each bracket are from $\psi 1$, $\psi 2$, $\psi A$ and $\psi F$ terms, respectively.   The dot lines denote terms which are too complex to be written in  (6.8).\\

$\bullet $ We now attempt to collect above results into the following form
\be  [\delta_{1},\delta_{2}]\psi_{a}=\upsilon^{\lambda}D_{\lambda}\psi_{ a}+\Lambda^{b}_{~a}\psi_{b}\ee   
to ensure the supersymmetry transformation to close into a translation and a gauge transformation.  Then, we find the following results:
\\

   1. We use $\kappa={2\over3}$ in  (5.14) to do following calculations.  

  2. The first bracket shall become $-{8\over4}D^{\mu}\psi_{a}$ to fit (6.9).  Thus 
\be
 -{3\over4}D^{\mu}\psi_{a} &=&{21\over4} \Gamma_{\nu\lambda} Q^{\mu\nu\lambda}\Gamma^{JK}\psi_b \phi^{J}_{c}\phi^{K}_{d} f^{bcd}_{~~~a}\nn\\
&~&+{3\over2}\Gamma^I \Gamma^{\nu\lambda}\Gamma_{\nu\bar\mu\bar\nu\bar\lambda} Q^{\bar\mu\bar\nu\bar\lambda}(\psi_{\bar d}D^\mu\phi^I_{\bar c}+\phi^I_{\bar c} D^\mu\psi_{\bar d} )G_{d\lambda} f^{\bar c\bar d}_{~~bc}f^{bcd}_{~~~a}
\ee 
This is  the  second constrain field equation (the first one is (5.16)) which relates $(\phi D\psi+\psi D\phi)$ to $\phi\phi\psi$ and $D\psi$.

  3.  Let second bracket =0 we then obtain the following fermion field equation
\be
{5\over8}\Gamma^{\mu}D_{\mu}\psi_{a} &-&{1\over8}  \Gamma_{\mu\nu\lambda} Q^{\mu\nu\lambda}\Gamma^{JK}\psi_b \phi^{J}_{c}\phi^{K}_{d} f^{bcd}_{~~~a}\nn\\
&-&{1\over4} \Gamma^{\mu\nu\lambda}\Gamma_{\nu\bar\nu\bar\lambda\bar\rho} Q^{\bar\mu\bar\nu\bar\lambda}\Gamma^I(\psi_{\bar d}D_\mu\phi^I_{\bar c}+\phi^I_{\bar c} D_\mu\psi_{\bar d} )G_{d\lambda} f^{\bar c\bar d}_{~~bc}f^{bcd}_{~~~a}=0
\ee
We can from (6.10) and (6.11)  find a simple form of the fermion field equation
\be
\Gamma^{\mu}D_{\mu}\psi_{a}+\Gamma_{\mu\nu\lambda} Q^{\mu\nu\lambda}\Gamma^{JK}\psi_b \phi^{J}_{c}\phi^{K}_{d} f^{bcd}_{~~~a}=0
\ee
and constrain field equation becomes
\be
\Gamma^I \Gamma^{\nu\lambda}\Gamma_{\nu\bar\mu\bar\nu\bar\lambda} Q^{\bar\mu\bar\nu\bar\lambda}(\psi_{\bar d}D^\mu\phi^I_{\bar c}+\phi^I_{\bar c} D^\mu\psi_{\bar d} )G_{d\lambda} f^{\bar c\bar d}_{~~bc}f^{bcd}_{~~~a}
+3 \Gamma_{\nu\lambda} Q^{\mu\nu\lambda}\Gamma^{JK}\psi_b \phi^{J}_{c}\phi^{K}_{d} f^{bcd}_{~~~a}=0
\ee 

   4. We can take the supervariation of  the fermion field equation to find the bosonic field equation, with the help of gauge field equation (5.15). 

  5.  To fit (6.9)  the fifth bracket term shall become $\Lambda_a^b\psi_{b}$ and  the other bracket terms ought to be zero.   This will lead to some additive constrain field equations.  Maybe theses extra constrain field equations automatically be satisfied after using the field equations.  Unfortunately, we have not yet found an elegant form for the final results and leave it aside for this moment.   

  6.  As that in [6] our $G_{a\mu}$ field is a super-invariant field and we also have a simple relation
\be
[\delta_1,\delta_2] G_{a\mu} =\upsilon^{\nu}D_{\nu}G_{a\mu}+\Lambda^{b}_{~a}G_{b\mu}~\Rightarrow~D_{\nu}G_{a\mu}=0,~~~\phi_b^I \phi_c^J G_{d\mu} f^{bcd}_{~~~a}=0
\ee
for the on-shell field.
\\

Let us finally remark that we should not overlook the self-duality property in M5-branes.  In the abelian case we need the field strength H to be anti-self-dual in order to close the algebra,  however, in this note we have not completed all the closures condition and we are not yet in the situation to concern the self-duality property in the loop space. (But we should note that perhaps we do not need the self-duality constraint to close the algebra in the loop space [7])

\section{Summary and Conclusion}
 In this note we start to study the relation between the (noncommutative) loop space, the Lie 3-algebra and multiple M5 branes. We use loop fields as the basic objects, the covariant derivative is given by the pull-back of the two-form potential and the gauge symmetry is described by the Lie 3-algebra which inevitably appear when we consider supersymmetry transformation. 

Guiding by the dimensional analysis, chiral property,  Fierz identity,  multiple D4 and single M5 supertransform properties, our SUSY transformation ansatz of non-abelian (2,0) theory is
\\
\be  \delta\phi^{I}_{a}&=&i\bar\epsilon \Gamma^{I}\psi_{ a}\\
\delta\psi_{a}&=&\Gamma^{\mu}\Gamma^{I}D_{\mu}\phi^{I}_{a} \epsilon
+\kappa\Gamma^{IJK}\Gamma^{\mu\nu\lambda} Q_{\mu\nu\lambda}\phi^{I}_{b}\phi^{J}_{c}\phi^{K}_{ d} f^{cdb}_{~~~a} \epsilon
+ \Gamma^{\mu\nu\lambda}F_{\mu\nu bc}G_{d\lambda}f^{bcd}_{~~~a}\epsilon\\
\delta A ^{b}_{\mu a}&=&i\bar\epsilon\Gamma_{\mu\nu\lambda\sigma}\Gamma^{I}Q^{\nu\lambda\sigma}\psi_{c}\phi^{I}_{d}f^{cdb}_{~~~a}\\  
\delta G _{a\mu}&=&0\ee 
We have examined the closure condition of the transformation and find 
\be \kappa = {2\over3}\ee 
The field equations are
\be 
 0&=& F_{\mu\nu a}^b-\varepsilon_{\mu\nu\lambda\rho\delta\sigma}\Big(\phi_c^I D^\sigma \phi_d^I + i {5\over 16}\bar \psi_{c}\Gamma^{\sigma}\psi_{d}\Big) Q^{\rho\lambda\delta}f^{cdb}_{~~~a}\\
 0&=&\Gamma^{\mu}D_{\mu}\psi_{a}+\Gamma_{\mu\nu\lambda} Q^{\mu\nu\lambda}\Gamma^{JK}\psi_b \phi^{J}_{c}\phi^{K}_{d} f^{bcd}_{~~~a}\\
0&=& D_\nu G_{a\mu}
\ee
The supervariation of  the fermion field equation can be used to  find the bosonic field equation, with the help of gauge field equation. 

The constrain field  equations we have found are 
\be
0&=&12 ( \epsilon_{\mu\nu\lambda}^{~~~~\bar\nu\bar\lambda\sigma}Q^{\nu\lambda\rho} F_{\rho\bar\nu\bar b\bar c}G_{\bar d\bar\lambda}+ 2  Q^{\nu\lambda\sigma} F_{\mu\nu\bar b\bar c}G_{\bar d\lambda})\phi_c^I~f^{\bar b\bar c\bar d}_{~~~d}\nn\\
&&-{3\over 8}\varepsilon_{\mu\nu\lambda\rho\delta}^{~~~~~~\sigma}Q^{\nu\lambda\rho}(\bar \psi_{c}\Gamma^{\delta}\Gamma^{I} \psi_{d})\\
0&=&\Gamma^I \Gamma^{\nu\lambda}\Gamma_{\nu\bar\mu\bar\nu\bar\lambda} Q^{\bar\mu\bar\nu\bar\lambda}(\psi_{\bar d}D^\mu\phi^I_{\bar c}+\phi^I_{\bar c} D^\mu\psi_{\bar d} )G_{d\lambda} f^{\bar c\bar d}_{~~bc}f^{bcd}_{~~~a}\nn\\
&&+3 \Gamma_{\nu\lambda} Q^{\mu\nu\lambda}\Gamma^{JK}\psi_b \phi^{J}_{c}\phi^{K}_{d} f^{bcd}_{~~~a}\\
0&=&\phi_b^I \phi_c^J G_{d\mu} f^{bcd}_{~~~a}
\ee
The closure conditions lead to several complex terms and we have not yet find the simple form for completed constrain field equations.  The detail of the closure conditions remain to be examined. 
\\

    However, a theory formulated in terms of non-local variables depending on loops is potentially very different from the usual formulations, we still have to make more investigations of this approach to see whether it will lead us into further insights of the mysterious multiple M5-branes.
\\
\\
\\
\\
\newpage
\begin{center}{\Large \bf APPENDIX} \end{center}
\begin{appendix}
\section{Symmetry Property and Chiral Property :} 
We work with 32-component Majorana spinors and the fermions are Goldstinos of the symmetry breaking SO(10, 1) $\rightarrow  SO(5, 1)\times SO(5)$ and $\mu,\nu,\lambda$ = 0, ..., 5; $I,J$ = 6, ..., 10.  Define
\be
\Gamma\equiv \Gamma ^{012345}
\ee
the convention for the chirality condition is therefore
\be  \Gamma \epsilon=+ \epsilon~~~~;\bar\epsilon \Gamma= - \bar\epsilon~~~~;~~~~\Gamma\psi=- \psi~~;~~\bar\psi \Gamma= \bar \psi\ee   
and the anti-commutative relations are
\be  \{ \Gamma_{\mu}, \Gamma_{I}\}=0~~;~~\{ \Gamma, \Gamma_{\mu}\}=0~~;~~[\Gamma, \Gamma_{I}]=0\ee   

 The conjugate spinors are defined with the charge conjugation matrix $C$
\be  
\bar \psi =  \psi^T C
\ee   
and for our representation we can choose $C= \Gamma^0$.  As $C^T= C^{-1}=-C$  we have
\be  
C\Gamma^\mu C^{-1}=- (\Gamma^\mu)^T\ee   
We thus find that
\be  
(C\Gamma_{a_1...a_n})^T=(\Gamma_{a_1...a_n})^T C^T=(-1)^{n-1}C\Gamma_{a_n...a_1}= (-1)^{n-1}(-1)^{[1+2+..+(n-1)]} C\Gamma_{a_1...a_n}
\ee   
and we have the following symmetry property about $\Gamma$  matrix
\be  
symmetry~matrix~& : &~~C\Gamma_{a_1},~C\Gamma_{a_1a_2},~C\Gamma_{a_1a_2a_3a_4a_5}~C\Gamma_{a_1a_2a_3a_4a_5a_6},...\\
anti-symmetry~matrix~& : &~~C\Gamma_{a_1a_2a_3},~C\Gamma_{a_1a_2a_3a_4},~C\Gamma_{a_1a_2a_3a_4a_5a_6a_7},...
\ee   
Usng above results we can find a simple rule of the sign under $ 1 \Leftrightarrow 2$ in below   :
\be  \bar\epsilon_{2}\gamma^{m}\epsilon_{1}&=&(-1)~\bar\epsilon_{1}\gamma^{m}\epsilon_{2}\nn  \\
\bar\epsilon_{2}\gamma^{mn}\epsilon_{1}&=& (-1)~\bar\epsilon_{1}\gamma^{mn}\epsilon_{2}\nn  \\
\bar\epsilon_{2}\gamma^{mno}\epsilon_{1}&=& (+1)~\bar\epsilon_{1}\gamma^{mno}\epsilon_{2}\nn  \\
\bar\epsilon_{2}\gamma^{mnop}\epsilon_{1}&=& (+1)~\bar\epsilon_{1}\gamma^{mnop}\epsilon_{2}\nn  \\
\bar\epsilon_{2}\gamma^{mnopq}\epsilon_{1} &=& (-1)~\bar\epsilon_{1}\gamma^{mnopq}\epsilon_{2}\ee   
From the chiral property $\Gamma\epsilon =\epsilon $  and $\bar \epsilon \Gamma = -\bar \epsilon$ we also see that
\be \bar \epsilon ~\Gamma^{\mu_1\cdot\cdot\cdot \mu_k}\Gamma^{I_1\cdot\cdot\cdot I_n}\epsilon&=& \bar \epsilon ~\Gamma^{\mu_1\cdot\cdot\cdot \mu_k}\Gamma^{I_1\cdot\cdot\cdot I_n}\Gamma\epsilon=(-1)^k~\bar \epsilon ~\Gamma\Gamma^{\mu_1\cdot\cdot\cdot \mu_k}\Gamma^{I_1\cdot\cdot\cdot I_n}~\epsilon\nn\\
&=& (-1)^{k+1}~\bar \epsilon ~\Gamma^{\mu_1\cdot\cdot\cdot \mu_k}\Gamma^{I_1\cdot\cdot\cdot I_n}~\epsilon
\ee
Thus, only if k is odd could we have no-zero value of $\bar \epsilon ~\Gamma^{\mu_1\cdot\cdot\cdot \mu_k}\Gamma^{I_1\cdot\cdot\cdot I_n}~\epsilon$.  Above symmetry property and chiral property  are used to derive the following Fierz identities and simplify several calculations in section 4-6.
\section{The Fierz identities :}
In 11 D spacetime the basis $\{Q^I\}$ of the vector space of $2^{11/2}\times 2^{11/2}$ matrices are
\be  
\{Q^b\}=\{ 1, \Gamma_m, i \Gamma_{mn}, i \Gamma_{mnp}, \Gamma_{mnpq}, \Gamma_{mnpqr} \}
\ee   
which satisfies the condition
\be
tr\{Q^aQ^b\}=2^{11/2}\delta_{ab}
\ee
From the basis in 11D we have the following general expansion
\be
(\bar \epsilon_2 \chi)\epsilon_1 &=& - 2^{-[\frac{11}{2}]} \Big((\bar \epsilon_2 \epsilon_1) \chi+(\bar \epsilon_2 \Gamma_{m}\epsilon_1)\Gamma^{m} \chi-\frac{1}{2!}(\bar \epsilon_2 \Gamma_{mn}\epsilon_1)\Gamma^{mn} \chi-\frac{1}{3!}(\bar \epsilon_2 \Gamma_{mnp}\epsilon_1)\Gamma^{mnp} \chi \nn  \\
&&+\frac{1}{4!}(\bar \epsilon_2 \Gamma_{mnpq}\epsilon_1)\Gamma^{mnpq} \chi+\frac{1}{5!}(\bar \epsilon_2 \Gamma_{mnpqr}\epsilon_1)\Gamma^{mnpqr} \chi\Big)
\ee 
Use the symmetry property in (A.9) we find the following combination in eleven-dimensions 
\be
(\bar \epsilon_2 \chi)\epsilon_1 -(\bar \epsilon_1 \chi)\epsilon_2 =-\frac{1}{16}\Big((\bar \epsilon_2 \Gamma_m\epsilon_1)\Gamma^m \chi-\frac{1}{2!}(\bar \epsilon_2 \Gamma_{mn}\epsilon_1)\Gamma^{mn} \chi+ \frac{1}{5!}(\bar \epsilon_2 \Gamma_{mnpqr}\epsilon_1)\Gamma^{mnpqr} \chi \Big)
\ee 
Consider the chirality property in (A.10) one then gets
\be
(\bar \epsilon_2 \chi)\epsilon_1 -(\bar \epsilon_1 \chi)\epsilon_2 &=&-\frac{1}{16}\Big((\bar \epsilon_2 \Gamma_\mu\epsilon_1)\Gamma^\mu \chi-(\bar \epsilon_2 \Gamma_{\mu}\Gamma^I\epsilon_1)\Gamma^\mu \Gamma^I \chi+ \frac{1}{3!}\frac{1}{2!}(\bar \epsilon_2 \Gamma_{\mu\nu\lambda} \Gamma^{IJ}\epsilon_1)\Gamma^{\mu\nu\lambda}\Gamma^{IJ} \chi \cr
&& + \frac{1}{4!}(\bar \epsilon_2 \Gamma_\mu \Gamma^{IJKL}\epsilon_1)\Gamma^\mu\Gamma^{IJKL} \chi + \frac{1}{5!}(\bar \epsilon_2 \Gamma_{\mu\nu\lambda\rho\sigma} \epsilon_1)\Gamma^{\mu\nu\lambda\rho\sigma} \chi \Big)
\ee
Translate the last line above in terms of fewer $\Gamma$-matrices with the help of $\epsilon$-tensors the final form of Fierz identity becomes [7,6] 
\be  (\bar\epsilon_{2}\chi)\epsilon_{1}-(\bar\epsilon_{1}\chi)\epsilon_{2}=-{1\over 16}\Big[~2(\bar\epsilon_{2}\Gamma_{\mu}\epsilon_{1})\Gamma^{\mu}\chi-2(\bar\epsilon_{2}\Gamma_{\mu}\Gamma^{I}\epsilon_{1})\Gamma^{\mu}\Gamma^{I}\chi\nn  +{1\over 12}(\bar\epsilon_{2}\Gamma_{\mu\nu\lambda}\Gamma^{IJ}\epsilon_{1})\Gamma^{\mu\nu\lambda}\Gamma^{IJ}\chi\Big]\nn\\
\ee 
We have used the above Fierz identity several times during calculating  the closures of supersymmetry transformation  in sections 4-6.
\section{The $\Gamma$ matrix (commutating property) :}
First, we can derive some simple relations 
\be  
 \Gamma^a\Gamma^{b} &=&{1\over2}(\Gamma^a\Gamma^b - \Gamma^b\Gamma^a)+ {1\over2}(\Gamma^a\Gamma^b + \Gamma^b\Gamma^a)\equiv\Gamma^{ab}+ g^{ab}\\
 \Gamma^a\Gamma^{bc} &=& {1\over2} \Gamma^a(\Gamma^b\Gamma^c - \Gamma^c\Gamma^b)={1\over2}(2g^{ab}-\Gamma^b\Gamma^a)\Gamma^c-{1\over2}(2g^{ac}-\Gamma^c\Gamma^a)\Gamma^b\nn\\
&=&g^{ab}\Gamma^c-{1\over2}\Gamma^b(2g^{ac}-\Gamma^c\Gamma^a)-g^{ac}\Gamma^b+{1\over2}\Gamma^c(2g^{ab}-\Gamma^b\Gamma^a)\nn\\
&=& \Gamma^{bc}\Gamma^a +2 g^{ab}\Gamma^{c}- 2 g^{ac}\Gamma^{b} \equiv  \Gamma^{bc}\Gamma^a+4 g^{a[b}\Gamma^{c]}\\
\Gamma^a\Gamma^{bcd} &=& -\Gamma^{bcd}\Gamma^a + g^{ab}(\Gamma^{cd}-\Gamma^{dc})+ g^{ad}(\Gamma^{bc}-\Gamma^{cb})+ g^{ac}(\Gamma^{db}-\Gamma^{bd})\nn  \\
 &\equiv&  -\Gamma^{bcd}\Gamma^a+6 g^{a[b}\Gamma^{cd]}
\ee   %
which will be used in sections  4-6.  These relations could be read from the formula
\be  [\Gamma^{a},\Gamma^{b_{1}.....b_{n}}]&=&(1-(-1)^{n})\Gamma^{ab_{1}.....b_{n}}+n(1+(-1)^{n})g^{a[b_{1}}\Gamma^{b_{2}.....b_{n}]}\\
\{\Gamma^{a},\Gamma^{b_{1}.....b_{n}}\}&=&(1+(-1)^{n})\Gamma^{ab_{1}.....b_{n}}+n(1-(-1)^{n})g^{a[b_{1}}\Gamma^{b_{2}.....b_{n}]}\ee   %
which could be derived from the relations
\be  \Gamma^{a}\Gamma^{b_{1}.....b_{n}}&=&\Gamma^{ab_{1}.....b_{n}}+ng^{a[b_{1}}\Gamma^{b_{2}.....b_{n}]}\\
\Gamma^{b_{1}.....b_{n}}\Gamma^{a}&=&\Gamma^{b_{1}.....b_{n}a}+n\Gamma^{[b_{1}.....b_{n-1}}g^{b_{n}]a}\ee   %

The most general formula is 
\be  \Gamma^{b_{1}.....b_{n}} \Gamma_{a_{1}....a_{n}}
=\sum_{p=0}^{min(n,m)} {n!m!\over (n-p)!(m-p)!p!}\Gamma^{[b_{1}.....b_{n-p}}~_{[a_{p+1}....a_{m}}g^{b_{n-p+1}.....b_{n}]}~_{{a_{1}....a_{p}}]}\\
\Gamma_{b_{1}.....b_{n}} \Gamma^{a_{1}....a_{n}}
=\sum_{p=0}^{min(n,m)} {n!m!\over (n-p)!(m-p)!p!}\Gamma_{[b_{1}.....b_{n-p}}~^{[a_{p+1}....a_{m}}g_{b_{n-p+1}.....b_{n}]}~^{{a_{1}....a_{p}}]}
\ee   %
For example 
\be  \Gamma_{\mu\nu\lambda\rho} \Gamma^{\bar\mu\bar\nu\bar\lambda}=\Gamma_{\mu\nu\lambda\rho}^{~~~~~\bar\mu\bar\nu\bar\lambda}
+12\Gamma_{[\mu\nu\lambda}^{~~~~[\bar\nu\bar\lambda}g_{\rho]}^{\bar\mu]}+36\Gamma_{[\mu\nu}^{~~~[\bar\lambda}g_{\lambda\rho]}^{\bar\mu\bar\nu]}+24\Gamma_{[\mu}g_{\nu\lambda\rho]}^{\bar\mu\bar\nu\bar\lambda]}
\ee   %
which will be used in section 6. 
\section{The $\Gamma$ matrix (summation property) :}
Using above relations we can find the following useful relations during the calculation of the closures (here a,b,c,...is the general index and n is defined by $\sum_a \Gamma^{a}\Gamma^{a} = n$)
\be  
\Gamma^{ab}\Gamma^{a}&=&\Gamma^{aba}+2\Gamma^{[a}g^{b]a}=0+(\Gamma^{a}g^{ba}-\Gamma^{b}g^{aa})=(1-n)\Gamma^{b}\\
\Gamma^{abc}\Gamma^{a}&=&\Gamma^{abca}+3\Gamma^{[ab}g^{c]a}=0+(\Gamma^{ab}g^{ca}+\Gamma^{ca}g^{ba}+\Gamma^{bc}g^{aa})=(n-2)\Gamma^{bc}\\
\Gamma^{abcd}\Gamma^{a}&=&\Gamma^{abcda}+4\Gamma^{[abc}g^{d]a}=0+(\Gamma^{abc}g^{da}+\Gamma^{dab}g^{ca}+\Gamma^{cda}g^{ba}+\Gamma^{bcd}g^{aa})\nn  \\
&=&(1+n)\Gamma^{bcd}\\
\Gamma^{de}\Gamma^{eabc}&=&(\Gamma^{d}\Gamma^{e}-g^{de})(\Gamma^{e}\Gamma^{abc}-3g^{e[a}\Gamma^{bc]})=(n-4)\Gamma^{d}\Gamma^{abc}-3g^{d[a}\Gamma^{bc]}
\ee   %
and
\be  \Gamma^{a}\Gamma^{b}\Gamma^{a}&=&(2g^{ab}-\Gamma^{b}\Gamma^{a})\Gamma^{a}= (2-n)\Gamma^{b}\\
\Gamma^{a}\Gamma^{bc}\Gamma^{a}&=& (\Gamma^{bc}\Gamma^a+4 g^{a[b}\Gamma^{c]})\Gamma^{a}=(n-4)\Gamma^{bc}\\
\Gamma^{a}\Gamma^{bcd}\Gamma^{a}&=& (-\Gamma^{bcd}\Gamma^a+6 g^{a[b}\Gamma^{cd]})\Gamma^{a}=(6-n)\Gamma^{bcd}
\ee   %
\
Also, the relations
 \be  
\Gamma^{abc}\Gamma^{d}\Gamma^{a}&=&\Gamma^{abc}(2g^{da}-\Gamma^{d}\Gamma^{a})=2\Gamma^{dbc}-(n-2)\Gamma^{bc}\Gamma^d\nn  \\
&=&(4-n)\Gamma^d\Gamma^{bc}-(12-4n)g^{d[b}\Gamma^{c]}\\
\Gamma^{abc}\Gamma^{de}\Gamma^{a}&=&\Gamma^{abc}(\Gamma^{a}\Gamma^{de}-4g^{a[d}\Gamma^{e]})=(n-2)\Gamma^{bc}\Gamma^{de}-2\Gamma^{abc}(g^{ad}\Gamma^e-g^{ae}\Gamma^d)\nn  \\
&=&(n-6)\Gamma^{de}\Gamma^{bc}
+(2n-10)[g^{dc}\Gamma^{be}-g^{ec}\Gamma^{bd}+g^{bd}\Gamma^{ec}-g^{be}\Gamma^{dc}]\nn  \\
&~&+4[g^{dc}g^{be}-g^{ec}g^{bd}]\ee   %
are used to investigate the Fermion field supersymmetry property in section 6.
\end{appendix}
\\
\\
\\
\begin{center} {\Large \bf REFERENCES}\end{center}
\begin{enumerate}
\item D. S. Berman, ``M-theory branes and their interactions," Phys. Rept. 456 (2008) 89 [arXiv:hep-th/0710.1707.]
\item  P. S. Howe, E. Sezgin, and P. C. West, ``Covariant field equations of the M-theory five-brane," Phys. Lett. B 399 (1997) 49V59, [arXiv:hep-th/9702008]\\
P. Pasti, D. P. Sorokin and M. Tonin, ``Covariant action for a D = 11 five-brane with the chiral field," Phys. Lett. B 398 (1997) 41  [arXiv:hep-th/9701037]\\
I. A. Bandos, K. Lechner, A. Nurmagambetov, P. Pasti, D. P. Sorokin and M. Tonin, ``Covariant action for the super-five-brane of M-theory," Phys. Rev. Lett. 78 (1997) 4332  [arXiv:hep-th/9701149]\\
M. Aganagic, J. Park, C. Popescu and J. H. Schwarz, ``World-volume action of the M-theory five-brane," Nucl. Phys. B 496 (1997) 191  [arXiv:hep-th/9701166]
\item  J. Bagger and N. Lambert, ``Modeling multiple M2," Phys. Rev. D 75 (2007) 045020  [arXiv:hep-th/0611108]\\
J. Bagger and N. Lambert, ``Gauge Symmetry and Supersymmetry of Multiple M2-Branes,"  Phys.Rev. D 77 (2008) 065008  [arXiv:hep-th/0711.0955 ] \\
J. Bagger and N. Lambert, ``Comments On Multiple M2-branes," JHEP 0802 (2008) 105  [arXiv:hep-th/0712.3738 ]\\
A. Gustavsson, ``Algebraic structures on parallel M2-branes," Nucl.Phys. B 811(2009) 66 [arXiv:hep-th/0709.1260]
\item  X. Bekaert, M. Henneaux, A. Sevrin, ``Chiral forms and their deformations," Commun. Math.  Phys. 224 (2001) 683 [arXiv:hep-th 0004049]
\item  E. Bergshoeff, ``The Mathematical Formulation of the M5-brane," Special Geometric Structures in string theory Bonn 8th-11th September, 2001\\
E. Bergshoeff, D. S. Berman, J. P. van der Schaar, P. Sundell, ``A Noncommutative M-Theory Five-brane," Nucl.Phys. B590 (2000) 173 [arXiv:hep-th/0005026]
\item N. Lambert and C. Papageorgakis, ``Nonabelian (2,0) Tensor Multiplets and 3-algebras," JHEP 1008 (2010) 083 [arXiv:hep-th1007.2982].
\item  A. Gustavsson,``The non-Abelian tensor multiplet in loop space," JHEP 0601 (2006)165 [arXiv:hep-th/0512341]
\end{enumerate}
\end{document}